\documentclass[twocolumn,showpacs,preprintnumbers,amsmath,amssymb]{revtex4}
\usepackage{dcolumn}
\usepackage{bm}
\begin{document}
\newcommand{\ba}{\begin{array}}
\newcommand{\ea}{\end{array}}
\newcommand{\be}{\begin{equation}}
\newcommand{\ee}{\end{equation}}
\newcommand{\bea}{\begin{eqnarray}}
\newcommand{\eea}{\end{eqnarray}}
\font\sqi=cmssq8
\def\DR{\rm I\kern-1.45pt\rm R}
\def\DC{\kern2pt {\hbox{\sqi I}}\kern-4.2pt\rm C}
\def\DH{\rm I\kern-1.5pt\rm H\kern-1.5pt\rm I}
\newcommand{\bs}{\mbox{\boldmath $\sigma$}}
\def\theequation{\arabic{equation}}
\title{Anisotropic  inharmonic  Higgs oscillator and related
(MICZ-)Kepler-like systems }
\author{Armen Nersessian$^{1,2}$ and  Vahagn Yeghikyan$^{1}$}
\affiliation{$\;^1$  Yerevan State University,  Alex Manoogian St. 1, Yerevan, 375025
    Armenia\\
 $\;^2$   Artsakh State University,  
 Stepanakert
   \& Yerevan Physics Institute, 
 Yerevan, 
    Armenia}
    \begin{abstract}
\noindent We propose the integrable (pseudo)spherical generalization of the
four-dimensional anisotropic oscillator with additional nonlinear potential.
 Performing its
Kustaanheimo-Stiefel transformation  we then  obtain the pseudospherical generalization of the MICZ-Kepler system with
linear and $\cos\theta$ potential terms. We also  present the generalization of the parabolic coordinates,
in which this system admits the separation of variables.
 Finally, we get the spherical analog  of  the presented MICZ-Kepler-like system.

\end{abstract}
\pacs{PACS numbers
14.80.Hv 02.30.Ik 03.65.-w 11.30.Pb}
\maketitle

\section{Introduction}
\noindent The oscillator and Kepler systems are the best known
examples  of mechanical systems with hidden symmetries \cite{perelomov}. Due to the existence of hidden symmetry
these systems admit separation of variables in few coordinate systems.
Despite of their qualitative difference, they can be related with
each other in some  cases.
Namely, $(p+1)-$dimensional Kepler system  can be obtained by the appropriate reduction procedures from the
$2p-$dimensional oscillator  for $p=1,2,4$ (for the review see, e.g. \cite{terant}).
These procedures, which  are known as
 Levi-Civita (or Bohlin) \cite{bohlin}, Kustaanheimo-Stiefel \cite{ks}
and Hurwitz \cite{h} transformations  imply
 the reduction of the oscillator  by the  action of $Z_2$,
$U(1)$, $SU(2)$ group, respectively, and yield, in general case, the Kepler-like
systems with  monopoles \cite{ntt,mic,su2}.
 The second system (with $U(1)$ (Dirac) monopole) is best
known and most important among them. It was invented independently by
Zwanziger and by McIntosh and Cisneros \cite{micz} and presently is refered
as  MICZ-Kepler system.

There are few
deformations of oscillator and Kepler systems, which preserve part
of hidden symmetries, e.g.,  anisotropic
oscillator,  Kepler  system with additional linear potential,
 two-center Kepler system \cite{perelomov},  as well as their ``MICZ-extensions" \cite{kno}.
 The Kepler system with linear potential is of  special importance due to its relevance to the Stark effect.
One can observe that the four-dimensional oscillator
 with additional anisotropic term
\be
U_A=\frac{\Delta\omega^2}{2}\sum_{i=1}^{p=2}(x^2_i-x^2_{i+p})
\label{delta}\ee
 results in the (MICZ-)Kepler system with potential
\be
 V_{cos}= \frac{\Delta\omega^2}{4}\cos\theta=\frac{\Delta\omega^2}{4}\frac{x_{p+1}}{|{\bf x}|},
  \label{cos}\ee
 which is the textbook example of the deformed Kepler system admitting the separation of variables
 in parabolic  coordinates. While (three-dimensional) Kepler system with additional linear potential
 (which is also separable in parabolic coordinates)
 is originated in the (four-dimensional) oscillator system
 with  fourth-order anisotropic potential term
\be U_{nlin}=-2 {\varepsilon}_{el}\sum_{i=1}^{p=2}
x^4_i-x^4_{i+p}. \label{nlin}\ee The corresponding potentials in
other dimensions look
 similarly.

Oscillator and Kepler systems admit the  generalizations on  a $d-$di\-men\-si\-onal
sphere and a two-sheet hyperboloid (pseudosphere). They are defined, respectively,
 by the following potentials \cite{sphere,sphere1}
\begin{equation}
U_{osc}=\frac{\omega^2 R_0^2}{2}\frac{{\bf x}^2}{{x}^2_{0}}, \quad V_{Kepler}=
-\frac{\gamma}{R_0}\frac{x_{0}}{|{\bf x}|},
\label{v}\end{equation}
where ${\bf x}, x_{0}$ are the Cartesian  coordinates of the ambient
(pseudo)Euclidean space $\DR^{d+1}$($\DR^{d.1}$):
$\epsilon{\bf x}^2+ x^2_{0}=R_0^2$, $\epsilon=\pm 1$.
 The $\epsilon=+1$ corresponds
to the sphere and  $\epsilon=-1$ corresponds to  the pseudosphere.
These systems also possess nonlinear hidden symmetries providing
them  with the properties similar to those
 of conventional oscillator and Kepler systems. Various aspects of these systems were
 investigated  in  \cite{2o}. Let us notice also mention the Ref. \cite{otchik},
  where the integrability of the  spherical two-center Kepler system was proved.

Completely similar to the planar case one can relate the oscillator and MICZ-Kepler
systems on pseudospheres (two-sheet hyperboloids). In the case of sphere, the relation between these systems is
slightly  different: the oscillator  on sphere results in the oscillator on hyperboloid
\cite{np}. After appropriate ``Wick rotation"  (compare with \cite{kmp}) of the MICZ-Kepler system on hyperboloid one can obtain the
MICZ-Kepler system on the sphere, constructed in \cite{kurochkin}.

As far as we know,  the integrable (pseudo)spherical analogs of the anisotropic oscillator  and of the oscillator with nonlinear
potential (\ref{nlin})
were unknown up to now,
as well as the (pseudo)spherical analog of  the (MICZ-)Kepler
system with linear and $\cos\theta$ potential terms. The construction
of these (pseudo)spherical systems is not only of  the academic interest.
They could be useful for the study of the various
physical phenomena
in nanostructures, as well as in the early  Universe. For example,
 the spherical generalization of the anisotropic oscillator potential
can be used as the confining potential  restricting the motion of particles
 in the asymmetric segments   of the thin (pseudo)spherical films.
 While with the (pseudo)spherical generalization of the linear potential at hands one can study the impact of the
 curvature of space in the Stark effect.

The construction of these systems is the goal of present  paper.
We shall present the integrable (pseudo)spherical
analog of four-dimensional oscillator  with the additional
anisotropic potentials (\ref{delta})  and (\ref{nlin}), given,
respectively, by the   expressions \be \frac{\Delta\omega^2}{2}
 {\bf x} \sigma_3 {\bf \bar x}\label{sadelta}\ee
and \be
\varepsilon_{el}R^2_0\frac{(R_0^2+x^2_0)}{x_0^4}({\bf x}{\bf\bar x}) ({\bf
x} \sigma_3 {\bf\bar x}), \label{inhasph}\ee where ${\bf
x}=x_\alpha +\imath x_{\alpha+2}$ are Cartesian coordinates of the
ambient (pseudo)Euclidean  space
  $\varepsilon {\bf x}{\bf \bar x}+ x_0^2=R^2_0$.

Then, performing Kustaanheimo-Steffel transformation, we get the
integrable Kepler system on pseudosphere with additional potential
terms generalizing   linear and $\cos\theta$ potentials of
ordinary (MICZ-)Kepler system.

These potentials can be written as follows \be \varepsilon_{el}
\frac{x_0}{R_0} x_3 +\frac{\Delta
\omega^2}{2}\left(\frac{x_3}{x}\pm x_0 x_3\right) .\label{arp}\ee
The upper sign corresponds to the potential reduced from
four-dimensional sphere, and lower sign corresponds to the one reduced
from the pseudosphere.
 We
present also the generalization of parabolic coordinates, where
the resulted system admits separation of variables. Finally,
performing ``Wick rotation" of the latter system we will obtain
the spherical analog of MICZ-Kepler system with linear and
$\cos\theta$ potentials: in terms of ambient space these potentials
are defined by the same expressions  as pseudospherical ones (\ref{arp}).

\section{Euclidean systems}\noindent
Let us start from the consideration of the Euclidean analog
of our construction. Namely, let us present the
 the integrable four-dimensional anisotropic inharmonic oscillator, and, performing
Kustaanheimo-Stiefel transformation, reduce it to  the MICZ-Kepler system with linear and $\cos\theta$ potentials.
It is convenient to describe the initial  four-dimensional system in complex coordinates
 \begin{equation}
 z^\alpha=\frac{x_1^\alpha +\imath x^\alpha_2}{\sqrt{2}},\quad
 \pi_\alpha=\frac{p_{1|\alpha} -\imath p_{2|\alpha}}{\sqrt{2}},
 \end{equation}
so that non-zero Poisson brackets between phase-space coordinates
look as follows
\begin{equation}
\{\pi_\alpha, z^\beta\}=1,\quad\{\bar\pi_\alpha, \bar z^\beta\}=1, \qquad \alpha,\beta=1,2 \; .
\label{pb}\end{equation}

In these coordinates the Hamiltonian of isotropic oscillator reads \be
{\cal H}_{0}=\pi\bar\pi+\omega^2 z\bar z. \label{plo}\ee
 Its rotational
symmetry generators are defined by the expressions \bea
&J=\frac{\imath}{2} (\pi z -\bar z\bar\pi ),&\label{j0}\\
&{\bf J}=\frac{\imath}{2} (\pi\bs z -\bar z\bs \bar\pi ),&\label{j3}\\
&J_{\alpha\beta}=\frac 12 \pi_\alpha\bar z^\beta ,\quad J_{\bar\alpha\bar\beta}=\frac 12 \bar\pi_\alpha z^\beta \;,&
\label{jg}\eea
and the hidden symmetry generators read
\bea
&{\bf A}=\frac 12 (\pi\bs\bar\pi +\omega^2 \bar z\bs z),&\label{dt}\\
&A_{\alpha\beta}=\frac 12 (\pi_\alpha\pi_\beta+\omega^2 \bar z^\alpha\bar z^\beta ),\quad
A_{\bar\alpha\bar\beta}={\overline  A_{\beta\alpha}}
\eea
The  integrable anisotropic inharmonic deformation of this system looks as follows
\bea
&{\cal H}_{aosc}={\cal H}_{0}+ ( \Delta\omega^2 + 2 {\varepsilon}_{el} z\bar z)z\sigma_3\bar z.\quad &\label{4a}
\eea
Its constants of motion are  given by (\ref{j0}), by  the third component of (\ref{j3}), and by the hidden symmetry generator
 \be A=
A_3+\frac{\Delta\omega^2}{2} (z\bar z)+
\frac{{\varepsilon}_{el}}{2}\left((z\bar z)^2+ (z\sigma_3\bar z)^2
\right), \label{4I}\ee  Clearly, the  potential term (\ref{delta})
decouples the  initial isotropic oscillator in the anisotropic one with the frequencies
$\omega_\pm=\sqrt{\omega^2\pm \Delta\omega^2}$. The second part of  the deformation term given by (\ref{nlin})
 has no such simple explanation. After transformation of the initial system  in the Kepler-like one
 it results in the linear potential.

{\bf Remark 1.} Assuming, that $z^\alpha$, are real coordinates   we arrive at the two-dimensional anisotropic inharmonic
oscillator. More generally, for $\alpha, \beta =1,\ldots N\geq 2$, and $\widehat\sigma_3$ is $N\times N$ dimensional
Hermitean matrix which obeys the condition $\widehat\sigma_3^2=1$ we get
 integrable anisotropic $4N-$($2N$-)dimensional inharmonic oscillator,
   when $z^\alpha$ are complex (real) coordinates. $\blacksquare$\\

Let us perform Kustaanheimo-Stiefel transformation of the presented system.
For this purpose we have to
reduce  the system under consideration by
the Hamiltonian action of the $U(1)$ group given by the generator
(\ref{j0})
and choose the $U(1)$-invariant  coordinates
\cite{ks,mic}
\begin{equation}
{\bf q}=z{\bs}{\bar z},\quad
{\bf p}=\frac{z{\bs}\pi+{\bar \pi}{\bs}{\bar z}}{2({z\bar z})},
\label{qp}\end{equation}
where $\bs$ are the Pauli matrices. \\
As a result,
the reduced Poisson brackets read
\begin{equation}
\{ p_i, q^j\}=\delta^j_i,\quad \{p_i, p_j\}=s\frac{\epsilon_{ijk}q^k}{q^3},\quad q=|{\bf q}|
\label{ss2}\end{equation}
where  $s$ is  value of the generator (\ref{j0}): $J=s$.
The oscillator's energy surface, ${\cal H}_{aosc}=E_{aosc}$ can be presented in the form
\begin{equation}
{\cal H}_{MICZ}={\cal E}_{MICZ}
\end{equation}
where \be {\cal H}_{MICZ} =\frac{{\bf p}^2}{2} +\frac{s^2}{2q^2}-
\frac{\gamma}{ q}+\frac{\Delta
\omega^2}{2}\frac{q_3}{q}+\varepsilon_{el}q_3,
 \label{C3}\ee and \be \gamma=\frac
{{E}_{aosc}}{2},\quad {\cal E}_{MICZ}=-\frac{\omega^2}{2} \ee
It is seen that (\ref{ss2}) and (\ref{C3})  define the MICZ-Kepler
system with the additional potential  (\ref{cos}) in the presence
of constant electric field  pointed along $x_3$-axes. For the completeness, let us
write down the constants of motion of the
constructed system reducing the constants of motion of the
four-dimensional oscillator. The $J_3$ results in the
corresponding component of angular  momentum,
 \be J={\bf n}_3{\bf J},\qquad {\bf J}={\bf p}\times{\bf
q} +s\frac{{\bf q}}{q}. \label{Jred}\ee
The reduced generator $A$ looks
as follows \be
 A={\bf n_3 A}+\frac{\varepsilon_{el}}{2}\left({\bf n}_3\times {\bf q}\right)^2 + \Delta\omega^2 \frac{\left({\bf n}_3\times {\bf q}\right)^2}{q}
\ee
where
\be
{\bf A}={\bf p}\times{\bf J}+\gamma\frac{{\bf q}}{q}
\ee
is the Runge-Lenz vector of the unperturbed MICZ-Kepler system.

Now, we are ready to consider similar oscillator-like systems on the four-dimensional sphere and pseudosphere,
 as well as the Kepler-like  systems on three-dimensional pseudosphere.

\section{Anisotropic inharmonic Higgs Oscillator}\noindent
For the description of the four-dimensional Higgs oscillator it is convenient to introduce
 the (complex) projective coordinates  connected with the Cartesian coordinates of five-dimensional ambient space
as follows
\begin{equation}
{\bf x}^\alpha\equiv x^\alpha +\imath x^{\alpha+2}=R_0\frac{2z^\alpha}{1+\epsilon z\bar z},\quad
 x_0=R_0\frac{1-\epsilon {z\bar z}}{1+\epsilon {z\bar z}}.
\label{x}\end{equation}
Here ${\bf x}{\bf\bar x}+\varepsilon x^2_0=R^2_0$, with $\varepsilon =1$ for the sphere and $\varepsilon=-1$
for the two-sheet hyperboloid.
In these coordinates the metric of  (pseudo)sphere reads
\begin{equation}
 ds^2=\frac{4R_0^2dz d\bar z}{(1+\epsilon{z\bar z})^2}.
\label{met}\end{equation}
where $|z|\in [0,\infty )$ for the sphere, and $|z|\in [0,1 )$ for the pseudosphere.
In the limit $R_0\to \infty$  the lower
 hemisphere (the lower sheet of hyperboloid) converts into the
 whole two-dimensional plane.

Now, defining the canonical Poisson brackets (\ref{pb}),
we can represent  the Hamiltonian of four-dimensional Higgs oscillator as follows
\begin{equation}
{ \cal H}^{\epsilon}_{0}= \frac{(1+\epsilon z{\bar
z})^2\pi{\bar\pi}}{2 R_0^2} +\frac{ 2 \omega^2R_0^2{z\bar
z}}{(1-\epsilon{z\bar z})^2}.
 \label{ho}\end{equation}
The    symmetries of (pseudo)sphere are defined by the generators
(\ref{j0})-(\ref{jg}), and
\begin{equation}
J_{\alpha}=(1-\epsilon z \bar z)\pi_\alpha+\varepsilon (\pi
z+\bar\pi\bar z){\bar z}^{\alpha}, \qquad J_{\bar\alpha}={\bar
J}_\alpha . \label{Jtr}\end{equation} It is clear that the generators
(\ref{j0})-(\ref{jg}) define the $so(4)$ (rotational) symmetry
algebra of the Higgs oscillator,
 while the generators (\ref{Jtr}) define the translations on (pseudo)the sphere.
 By their use
 one can construct
 the generators of hidden symmetries of the Higgs oscillator,
\begin{equation}
{ A}_{\alpha\beta}=\frac{{J}_\alpha{ J}_\beta}{2 R_0^2} +2
\omega^2 R_0^2\frac{{\bar z}^\alpha{\bar z}^\beta}{(1-\epsilon z\bar
z)^2},\quad I_{\bar\alpha\bar\beta}= {\bar I}_{\alpha\beta}
\label{Ic}\end{equation} and
\begin{equation}
{\bf A}=\frac{({ J}\bs{\bar J})}{2 R_0^2} + 2
\omega^2R_0^2\frac{(z\bs{\bar z})}{(1-\epsilon z\bar z)^2}.
\label{Ics}\end{equation}\\

Let us construct the integrable (pseudo)spherical analog of
the anisotropic inharmonic oscillator (\ref{4a}). We consider the class of Hamiltonians
 \be {\cal
H}^{\epsilon}_{aosc}={\cal H}^\epsilon_{0}+ (z\sigma_3\bar
z) \Lambda (z\bar z),
 \label{HA}\ee
 which  besides the symmetries defined by the generators $J$ and $J_3$, possess the hidden symmetry
 defined by the constant of motion
\be A=A_3 +  g(z\bar z) +(z\widehat\sigma_3\bar z)^2 h(z\bar
z). \ee
 Here $\Lambda (z\bar z)$,  $g(z\bar z)$ and $h(z\bar z)$ are  some  unknown
functions, and $A_3$ is the third component of (\ref{Ics}).

Surprisingly, from the requirement that $A$ is the constant of motion,
we uniquely (up to constant parameters) define the functions $\Lambda, g, h$, i.e. find the integrable anisotropic
generalization  of Higgs oscillator.
 Namely, the function $\Lambda$ in (\ref{HA}) reads
 \be {\Lambda}\equiv \frac{2 R^2_0\Delta\omega^2
}{(1+\epsilon z\bar z )^2}+ \frac{8 \varepsilon_{el}
R^4_0}{(1-(z\bar z)^2)^2}
 \frac{(1+(z\bar z)^2)(z \bar z) }{(1-\epsilon z\bar z )^2},
\label{R}\end{equation}
and the hidden symmetry generator looks as follows
$$
A={A}_3 + \frac{2 R^2_0\Delta\omega^2{z\bar z}}{(1+\epsilon z\bar
z )^2}+
$$
\begin{equation}+ 4\varepsilon_{el} R^4_0 \left(
 \frac{({z\bar z})^2}{(1- (z\bar z)^2 )^2}+
\frac{({z \widehat\sigma_3\bar z})^2}{(1-\epsilon z\bar z
)^4}\right). \label{IA}\end{equation}
One can easily see that the constructed system results in (\ref{4a}) results in the limit $R_0\to \infty$.

{\sl  Hence,
we have got the well-defined (pseudo)spherical generalization of
(\ref{4a})}.

In coordinates (\ref{x})
the potential of the constructed system looks much simpler.
The potential of (isotropic) Higgs reads \be
U_{Higgs}=\frac{\omega^2R_0^2}{2}\frac{R_0^2-x_0^2}{x_0^2}, \ee
while the anisotropy terms is defined by the expression \be
U_{AI}=\left(\frac{\Delta \omega^2}{2} +
\epsilon\varepsilon_{el}R^2_0\frac{(R_0^4-x_0^4)}{x_0^4}\right) {\bf x} \widehat\sigma_3 {\bf \bar x}
\label{arp2}\end{equation}
\section{ MICZ-Kepler-like systems on pseudosphere}\noindent
In this Section performing Kustaanheimo-Stiefel transformation
of the constructed system we shall get the pseudospherical analog of  the Hamiltonian (\ref{C3}).
This procedure  is completely similar to those of the isotropic Higgs oscillator
\cite{np}.

At first,  we must reduce the system by the Hamiltonian action  of the generator (\ref{j0}). Choosing the functions (\ref{qp})
as  the reduced coordinates, and fixing the level surface $J=s$, we shall get the six-dimensional
 phase space equipped by the Poisson brackets (\ref{ss2}).
 Then we fix the energy surface of the oscillator on the (pseudo)sphere,
${\cal H}^\epsilon_{aosc}=E_{aosc}$, and multiply it by $(1-\epsilon q^2)^2/q^2$. As a result,  the  energy surface
of the reduced  system takes the form
\begin{equation}
{\cal H}^{-}_{AMICZ}={\cal E}^{-}_{AMICZ},
\end{equation}
where
$$
{\cal H}^{-}_{AMICZ}= \frac{(1- { q}^2)^2}{8r_0^2}({\bf p}^2
+\frac{s^2}{{q}^2})- \frac{\gamma}{2r_0}\frac{1+{ q}^2}{q}+$$ \be
+ \frac{\Delta\omega^2}{2}\left(\frac{1-\epsilon {q}}{1+\epsilon
{q}}\right)^2 \frac{q_3}{q}+2 \varepsilon_{el}
r_0\frac{1+{q^2}}{1-{q^2}} \frac{q_3}{1-{q^2}},
\label{C31}\end{equation}
\begin{equation}
r_0=R_0^2,\quad  \gamma= \frac{E_{aosc}}{2}, \quad {\cal E}^{-}_{AMICZ}=-
\frac{\omega^2}{2}+\epsilon\frac{E_{aosc}}{2 r_0}.
\label{gam}\end{equation} Interpreting ${\bf q}$  as the
 stereographic coordinates of three-dimensional
pseudosphere
\begin{equation}
{\bf x}=r_0\frac{2{\bf q}}{1- {q}^2},\quad
 x_0=r_0\frac{1+ {q}^2}{1- { q}^2},
\label{hyper3}
\end{equation}
  we conclude that (\ref{C31}) defines
    the
    pseudospherical analog of the
MICZ-Kepler system with linear and $\cos\theta$ potential terms (\ref{C3}).

The constants of motion of the anisotropic oscillators, $J_3$ and
${A}$ yield, respectively, the third component of angular momentum
(\ref{Jred}) and the  hidden symmetry generator \be A={\bf n}_3
{\bf A}+\frac{ r_0 \Delta \omega^2}{ (1+\epsilon q)^2} \left[\frac
{q^2-q_3^2}{q}\right]+2 \varepsilon_{el} r^2_0
\frac{q^2-q_3^2}{(1-q^2)^2} \ee where
$$
 {\bf A}=\frac{{\bf T}\times{\bf  J}}{2 r_0}+
\gamma \frac{{\bf q}}{q}
$$
is the Runge -Lenz vector  of the MICZ-Kepler system on pseudosphere,  ${\bf J}$ is the  generator of the rotational momentum
defined by the expression (\ref{Jred}),
and
\begin{equation}
{\bf T}=\left(1+{q^2}\right){\bf p}-2 ({\bf qp})\;{\bf q}.
\end{equation}
is translation generator.

This term also looks simply in Euclidean coordinates of ambient
space: \be V_{AI}=\frac{\Delta
\omega^2}{2}\left(\frac{x_3}{x}+\epsilon x_0
x_3\right)+\varepsilon_{el}x_0 x_3 \ee
Let us notice, that the term proportional to $\Delta\omega^2$ depends on $\epsilon$, i.e., formally, the anisotropic
terms yield different pseudospherical
 generalizations of potential (\ref{cos}). However, this difference is rather trivial: it is easy to observe, that
 one potential transforms
 in other one upon spatial reflection.

 Presented
 Kepler-like system
admits the separation of variables  in the following generalization of parabolic coordinates (compare with  \cite{bogush}):
$$
q_1 +iq_2 = \frac{2{\sqrt {\xi \eta } }}{{r_0 + \frac{{\sqrt
{\sqrt{(r^2_0 + \xi ^2 )(r^2_0 + \eta ^2 )} + \xi \eta  + r^2_0}
}}{{\sqrt 2 }}}}{\rm e}^{\imath\varphi},$$
\begin{equation}
q_3=\frac{\sqrt{2}{\sqrt {\sqrt{(r^2_0 + \xi ^2 )(r^2_0 + \eta ^2
)} - \xi \eta - r^2_0} }}{r_0+\frac{{\sqrt {\sqrt{(r^2_0 + \xi ^2
)(r^2_0 + \eta ^2 )} + \xi \eta  + r^2_0} }}{{\sqrt 2 }}}.
\label{pc}\end{equation} In these coordinates the metric reads
$$
ds^2= $$
\begin{equation}
=r_0^2\frac{\xi+\eta }{4}\left(
\frac{d\xi^2}{\xi(r_0^2+\xi^2)}+\frac{d\eta^2}{\eta(r^2_0+\eta^2)}\right)+
\xi\eta d\varphi^2.
\end{equation}
Passing  to the canonical momenta, one can represent the Hamiltonian (\ref{C31}) as follows
$$
{\cal H}^-_{MICZ}=\frac{2 {\xi (r_0^2+\xi^2)}}{{r^2_0(\xi+\eta)
}}p_\xi^2+\frac{2 {\eta (r_0^2+\eta^2)}}{{r^2_0(\xi+\eta)
}}p_\eta^2+\frac{1}{\xi\eta}\frac{p_{\varphi}^2}{2}+
$$
$$
\frac{s
p_{\varphi}+s^2}{r_0(\xi+\eta)}\left(\frac{r_0+\sqrt{r_0^2+\xi^2}}{\xi}+\frac{r_0-\sqrt{r_0^2+\eta^2}}{\eta}\right)+
$$
$$+\frac{\Delta\omega^2 r_0}{2 }\frac{\xi
\sqrt{r_0^2+\xi^2}-\eta\sqrt{r^2_0+\eta^2}+\xi^2-\eta^2}{\xi+\eta}-
$$
\begin{equation}
-\frac{\gamma}{r_0}
\frac{\sqrt{r_0^2+\xi^2}+\sqrt{r^2_0+\eta^2}}{\xi+\eta}+\varepsilon_{el}\frac{\xi-\eta}{2}
\end{equation}
So, the corresponding generating function has to have the additive form-
$S={\cal E}_{AMICZ}t+p_{\varphi}\varphi+S_1(\xi)+S_2(\eta)$ . Replacing
$p_{\xi}$ and $p_{\eta}$ by ${dS_1(\xi)}/{d{\xi}}$ and
${dS_2(\eta)}/{d{\eta}}$ respectively, we obtain the following ordinary differential equations
$$
\frac{2\xi (r_0^2+\xi^2)}{r_0^2}
\left(\frac{dS_1(\xi)}{d{\xi}}\right)^2+(s
p_{\varphi}+s^2)\frac{r_0+\sqrt{r_0^2+\xi^2}}{r_0\xi}
$$
$$+\frac{\Delta\omega^2 r_0}{2 }(\xi
\sqrt{r^2_0+\xi^2}+\xi^2)-$$
\begin{equation}
-\frac{\gamma}{r_0}\sqrt{r^2_0+\xi^2}+\varepsilon_{el}
\xi^2 -{\cal E}_{AMICZ}\xi+\frac{p_{\varphi}^2}{\xi}=\beta
\end{equation}
$$
\frac{2\eta
(r^2_0+\eta^2)}{r^2_0}\left(\frac{dS_2(\eta)}{d{\eta}}\right)^2+(s
p_{\varphi}+s^2)\frac{r_0-\sqrt{r^2_0+\eta^2}}{r_0\eta}+ $$
$$
-\frac{\Delta\omega^2 r_0}{2 }(\eta \sqrt{r^2_0+\eta^2}+\eta^2)-
$$
\begin{equation}-\frac{\gamma}{r_0}\sqrt{r^2_0+\eta^2}-\varepsilon_{el} \eta^2
-{\cal E}_{AMICZ}\eta+\frac{p_{\varphi}^2}{ \eta}=-\beta
\end{equation}
From these equations we can immediately find the explicit expression for the generating function.
We have separated the variables for the pseudospherical generalization of the Coulomb system with linear and
$\cos\theta$ potential.

The above equations looks much simpler in the new coordinates $(\chi, \zeta)$, where
 $\xi=r_0 \sinh\chi $, $\eta=r_0 \sinh\zeta$.
$$
\left(\frac{dS_1(\chi)}{d\chi}\right)^2=\frac{{\cal E}_{AMICZ}}{2}
-\frac{\Delta\omega^2 r_0^4}{2}(\cosh\chi+\sinh \chi)+$$
$$
+(\frac{\gamma
r_0}{2}- s^2-sp_{\varphi})\coth\chi
-\frac{\varepsilon_{el} r_0^3}{2}
\sinh\chi -\frac{p_{\varphi}^2}{2 \sinh^2\chi}+$$
\begin{equation}
+\frac{\beta r_0 - s^2-sp_{\varphi}}{2
\sinh\chi},
\end{equation}
$$
\left(\frac{dS_2(\zeta)}{d\zeta}\right)^2= \frac{{\cal E}_{AMICZ}}{2}
+\frac{\Delta\omega^2
r_0^4}{2}(\cosh\zeta+\sinh\zeta)+$$
$$
+(\frac{\gamma r_0}{2} +s^2+sp_{\varphi})\coth\zeta
 +\frac{\varepsilon_{el}r_0^3}{2}
\sinh\zeta -\frac{p_{\varphi}^2}{2 \sinh^2\zeta}-
$$
\begin{equation}-
\frac{\beta
r_0+s^2+sp_{\varphi}}{2\sinh\zeta}.
\end{equation}

{\bf Remark 2.} In the same manner  the  $2p-$dimensional anisotropic inharmonic oscillator on (pseudo)sphere
can be  connected to the $(p+1)-$dimensional
Kepler-like systems on pseudosphere also for the  $p=1,4$ . For $p=1$ we should just assume that $z^\alpha$ are
{\sl real} coordinates. In this  case we should not perform any reduction at the classical level
(in quantum case we have to reduce the initial
 system by the discreet $Z_2$ group action, see \cite{ntt}).
 For the $p=4$ we have to assume, that $z^\alpha$ are
{\sl quaternionic} coordinates (equivalently,  that $z^\alpha $ are complex coordinates with $\alpha=1,\ldots 4$).
In contrast with $p=1,2$ cases, we should   reduce the initial system by the $SU(2)$ group
action \cite{su2}.$\blacksquare$\\

{\bf Remark 3.} The planar (MICZ)-Kepler system with linear potential can be obtained as a limiting case of the two-center
(MICZ-) Kepler system, when one of the forced centers is placed at infinity (see, e.g. \cite{perelomov}).
The two-center (pseudo)spherical Kepler system is the  integrable system as well \cite{otchik}. However, presented
 pseudospherical generalization of the (MICZ-)Kepler system with linear potential could not be obtained from  the
two-center pseudospherical Kepler system: it can be easily checked, that in contrast with pseudospherical Kepler potential,
it does not obey the corresponding  Laplas equation.
 $\blacksquare$

\section{Transition to the sphere}
\noindent
To get the spherical counterpart of the Hamiltonian (\ref{C3}),
let us perform its ``Wick rotation" which yields
$$
{\cal H}^+={\cal H}^+_{0}+2 \varepsilon_{el}r_0\frac{1-{q^2}}{1+{q^2}}
\frac{q_3}{1+{q^2}}+
$$
\begin{equation}
 +\frac{\Delta\omega^2}{2}
\left(\frac{1-i\epsilon {q}}{1+i\epsilon {q}}\right)^2
\frac{q_3}{q}, \label{C32}\end{equation} where \be {\cal
H}^+_{0}=\frac{(1+{q^2})^2}{8 r^2_0}\left({\bf
p}^2+\frac{s^2}{q^2}\right)^2-\gamma\frac{1-{q^2}}{2r_0 q} \ee is
the Hamiltonian of unperturbed MICZ-Kepler system on the sphere.
The hidden symmetry of this system is defined by the expression
\be A={\bf n}_3{\bf A}+ \Delta \omega^2\  \left[\frac
{q^2-q_3^2}{(1+i\epsilon q)^2 q}\right]+2 \varepsilon_{el}
r_0^2\frac{q^2-q_3^2}{(1+q^2)^2} .\label{Ac}\ee where \be {\bf A}={\bf
J}\times{\bf T}+\gamma \frac{{\bf q}}{q} \label{RL}\ee
is Runge-Lenz vector of the spherical MICZ-Kepler system, with the angular momentum ${\bf J}$ given by (\ref{Jred})
and with the
translation generator
\begin{equation}
{\bf T} = \left(1 -{q^2}\right){\bf p} +2({\bf qp}){\bf q}.
\end{equation}
One can see, that due to the last term in (\ref{C32})  this Hamiltonian is a complex one.
Taking its real part we shall get the integrable spherical analog of the
MICZ-Kepler system with linear and $\cos\theta $ potentials,
$$
{\cal  H}^{+}_{MICZ}={\cal H}^+_{0}
+\frac{\Delta\omega^2}{2}\frac{1-6q^2+q^4}{1+q^2}
\frac{q_3}{q}+$$\begin{equation}+2\varepsilon_{el}\frac{1-{q^2}}{1+{q^2}}
\frac{q_3}{1+{q^2}}.
\end{equation}
The generator of its hidden symmetry is also given by the real part of (\ref{Ac})
\begin{equation}
A={\bf n_3}{\bf A}+\left[\Delta \omega^2 r_0 \frac{1-{q^2}}{q}+\frac{\varepsilon_{el}}{2}\right]
\frac
{q^2-q_3^2}{ (1+{q^2})^2 }.
\end{equation}\\
In the terms of ambient space $\DR^{4}$ the anisotropy term  is defined by the expression (\ref{arp2}).

{\bf Remark 4.} It is clear from our
consideration, that the addition
 to the constructed system of the
 potential \be c_0\; {\rm
Im}\;\left(\frac{1-i\epsilon {q}}{1+i\epsilon {q}}\right)^2
\frac{q_3}{q}\label{va}\ee will also preserve the integrability. The hidden symmetry generator will be  given by the
expression \be A+c_0\; {\rm Im}\frac{\Delta \omega^2}{2
(1+i\epsilon q)^2} \left[\frac {q^2-q_3^2}{q}\right]. \ee However,
it is easy to see, that this additional potential coinsides with (\ref{C32}),
i.e. we do not get anything new in this way. $\blacksquare$
\section{Conclusion}\noindent
We presented  the integrable (pseudo)spherical generalization  of anisotropic oscillator which can be considered as
a deformation of the well-known
Higgs oscillator. By the use of this system we constructed the integrable (pseudo)spherical analog of the MICZ-Kepler system
with linear and $\cos\theta$ potentials.
We proved the integrability of these systems postponing the study of its classical
and quantum mechanical
solutions. The computation of the quantum mechanical spectrum of these systems, and, consequently,
the clarification of the impact of the space curvature
in the Stark effect is a problem of special interest especially interesting problem from the viewpoint of
the mesoscopic physics and of the cosmology, as well. Let us notice that even in the flat space the presence of the Dirac
monopole leads
to qualitative changes of the properties of Stark effect \cite{stark}. There is no doubt  that similar phenomena will
appear in the Stark effect on curved space.
Taking into account the conclusions of recent papers \cite{kno,no}, we expect that one can preserve the integrability of the
 proposed
system, introducing the constant magnetic field and the appropriate potential term.
 In that case we shall have at hands the   integrable system in the parallel ``homogeneous" electric and magnetic fields.
 The importance of such system is obvious.\\

{\large Acknowledgments.} The authors are thankful to Levon Mardoyan, Vadim Ohanyan  and
George Pogosyan for useful comments, and to Victor Redkov for providing  us with the reprint of the
paper \cite{otchik}.
The work is supported by the Artsakh Ministry of Education and Science and
by the    grants NFSAT-CRDF-ARP1-3228-YE-04 and NFSAT-CRDF-UC 06/07, INTAS-05-7928  grant.


\begin{thebibliography}{99}
\bibitem{perelomov}A.~M.~Perelomov,
{\sl Integrable systems
 of classical mechanics and Lie algebras}, Birkhauser, 1990
\bibitem{terant}V.~M.~Ter-Antonyan, {\sl  Dyon-Oscillator duality},
[quant-ph/0003106]

%
%
%
%
%


\bibitem{bohlin}T. Levi-Civita, Opere Mathematiche,{\bf 2} 411 (1906);

K.~Bohlin, Bull.~Astr., {\bf 28}  144 (1911)

\bibitem{ks}P.~Kustaanheimo,~E.~Stiefel,~J.~Reine Angew Math., {\bf 218}  204 (1965)
\bibitem{h}A.~Hurwitz,~Mathematische~Werke,~Band~II,~641
({\sl Birkh\"auser},~Basel,~1933)

L.~S.~Davtyan {\it et al}, J.~Phys.~{\bf A20}~6121~(1987);

D.~Lambert, M.~Kibler, J.~Phys.~{\bf A21},~307~(1988)
\bibitem{ntt}A.~Nersessian,~V.~M.~Ter-Antonyan,~M.~Tsulaia, Mod.~Phys.~Lett.
{\bf A11}~1605~(1996);
\bibitem{mic}T.~Iwai,~Y.~Uwano, J.~Math.~Phys.~{\bf 27},~1523~(1986)

             I.M. Mladenov and V.V. Tsanov, 
             {J. Phys.}, {\bf A20}, 5865 (1987);    { J. Phys.}, {\bf A32}, 3779 (1999).
\bibitem{su2}T.~Iwai,~J.~Geom.~Phys.~{\bf 7},~507~(1990);

L.~G.~Mardoyan,~A.~N.~Sissakian,~V.~M.~Ter-Antonyan,
Phys.~Atom.~Nucl.~{\bf 61},~1746~(1998)

       \bibitem{micz}   D.~Zwanziger,{Phys.~Rev.} {\bf 176},~1480 ~(1968)

       H. McIntosh and A. Cisneros, {J. Math. Phys.} {\bf 11}, 896
       (1970).


  \bibitem{kno}S.~Krivonos, A.~Nersessian and  V.~Ohanyan
 {\em Phys.\ Rev.\/} {\bf D75}, 085002 (2007)
\bibitem{no}A.~Nersessian and V.~Ohanyan,
 [arXiv:0705.0727],
 to appear in {\em Theor. Math. Phys\/}

\bibitem{sphere}E.~Schr\"odinger,~Proc.~Roy.~Irish~Soc.~{\bf 46}~9~(1941); {\bf 46}~183
(1941); {\bf 47}~53~(1941)
\bibitem{sphere1}P.~W.~Higgs,~J.~Phys.~A:~Math.~Gen.~{\bf 12}~309~(1979)

H.~I.~Leemon,~J.~Phys.~A:~Math.~Gen.~{\bf 12}~489~(1979)

\bibitem{2o}A.~Barut, A.~Inomata, G.~Junker,~J.~Phys. {\bf A20},~6271~(1987);
J.~Phys.~{\bf A23}~1179~(1990),

Ya.~A.~Granovsky,~A~.S.~Zhedanov,~I.~M.~Lutzenko,
Teor.~Mat.~Fiz.{\bf 91}~207~(1992); {\bf 91}~396~(1992);

D.~Bonatos,~C.~Daskaloyanis, K.~Kokkatos, Phys.~Rev. {\bf A50}~3700~(1994);

E.~G.~Kalnins,~W.~Miller~Jr.,~G.~S.~Pogosyan, J.~Math.~Phys.
{\bf 37}~6439~(1996);{\bf 38}~5416~(1997)

 A. Ballesteros, F. J. Herranz, O. Ragnisco,  J.Phys. {\bf A38}  7129 (2005);

A. Ballesteros, F. J. Herranz, J.Phys. {\bf A40} F51(2007)


J. F. Carinena, M. F. Ranada, M. Santander, Ann. Phys. {\bf 322},  2249 (2007)


\bibitem{otchik}V.~S.~Otchik, Doklady Akademii Nauk BSSR, {\bf 35}, No. 5, 420 (1991)(in Russian)
\bibitem{np} A.~Nersessian and G.~Pogosyan,
 {\em Phys.\ Rev.\/} {\bf A63}, 020103(R) (2001)

\bibitem{kmp}E.~G.~Kalnins,~W.~Miller~Jr.,~G.~S.~Pogosyan,
J.~Math.~Phys.{\bf 37} 2629 (2000)
 \bibitem{kurochkin}
V.~V.~Gritsev, Yu.~A.~Kurochkin, V.~S.~Otchik,
{\em J.~Phys.\/} {\bf A33}, 4903 (2000)


\bibitem{bogush} A.~A.~Bogush, V.~S.~Otchik and V.~M.~Red'kov,
  arXiv:hep-th/0612178.
%
%
\bibitem{stark}
      L.~Mardoyan {\sl et al},
       { Theor.\ Math.\ Phys.},  {\bf 140},  958 (2004);  
         { J.\ Phys.} {\bf A40}, 5973 (2007)  





\end{thebibliography}
\end{document}